# Resistivity phase diagram of cuprates revisited

D. Pelc[1,*], M. J. Veit[1,‡], C. J. Dorow[1,§], Y. Ge[1,‖], N. Barišić[2,3*] and M. Greven[1,*]

[1] School of Physics and Astronomy, University of Minnesota, Minneapolis, MN 55455, USA

[2] Institute of Solid State Physics, TU Wien, 1040 Vienna, Austria

[3] Department of Physics, Faculty of Science, University of Zagreb, HR–10000 Zagreb, Croatia

Present addresses:

[‡] Department of Applied Physics, Stanford University, Stanford, CA 94305, USA

[§] Department of Physics, University of California, San Diego, CA 92093, USA

[‖] Department of Physics, Penn State University, University Park, PA 16802, USA

* correspondence to: dpelc@umn.edu, barisic@ifp.tuwien.ac.at, greven@umn.edu

**The phase diagram of the cuprate superconductors has posed a formidable scientific challenge for more than three decades. This challenge is perhaps best exemplified by the need to understand the normal-state charge transport as the system evolves from Mott insulator to Fermi-liquid metal with doping. Here we report a detailed analysis of the temperature ($T$) and doping ($p$) dependence of the planar resistivity of simple-tetragonal $HgBa_2CuO_{4+\delta}$ (Hg1201), the single-$CuO_2$-layer cuprate with the highest optimal superconducting transition temperature, $T_c$. The data allow us to test a recently proposed phenomenological model for the cuprate phase diagram that combines a universal transport scattering rate with spatially inhomogeneous (de)localization of the Mott-localized hole. We find that the model provides an excellent description of the data. We then extend this analysis to prior transport results for several other cuprates, including the Hall number in the overdoped part of the phase diagram, and find little compound-to-compound variation in (de)localization gap scale. The results point to a robust, universal structural origin of the inherent gap inhomogeneity that is unrelated to doping-related disorder. They are inconsistent with the notion that much of the phase diagram is controlled by a quantum critical point, and instead indicate that the unusual electronic properties exhibited by the cuprates are fundamentally related to strong nonlinearities associated with subtle nanoscale inhomogeneity.**

## I. Introduction

The cuprates exhibit multiple electronic ordering tendencies, a partial depletion of states at the Fermi level (a 'pseudogap', PG), and unusual ('strange-metal') normal-state transport behavior [1]. One of the most extensively investigated observables is the planar resistivity, $\rho$, which features several well-established characteristics [2,3,4,5]: $T$-linear behavior above the doping-dependent PG temperature, $T^*(p)$, which is most pronounced near optimal doping, where $T^*$ approaches $T_c$ [3,4], and simple $T^2$ behavior at very high doping, where other measurements confirm Fermi-liquid (FL) quasiparticle behavior associated with a density of $1 + p$ carriers [2,6,7]. Based on systematic resistivity data for a particular cuprate compound, the normal-state phase diagram can be obtained by evaluating the curvature (i.e., the second temperature derivative) of $\rho(T,p)$ [3]. Resistivity measurements thus provide indispensable information about the cuprates, yet until recently, a comprehensive understanding of this pivotal observable had remained elusive.

In the theory of charge transport, the resistivity is generally parametrized with a charge-carrier density, effective mass, and transport scattering rate. In principle, these quantities may be temperature- and doping-dependent, and therefore additional insight is needed to understand the resistivity. Two common assumptions made in evaluating the properties of the cuprates are that the temperature dependence of the resistivity is solely determined by the temperature dependence of the effective scattering rate, and that the PG opens at the doping-dependent temperature $T^*(p)$. The former assumption neglects the depletion of the density of states due to the opening of the PG, and hence the possibility that the carrier density depends on temperature, whereas the latter assumption overlooks the experimental evidence that local gaps already form at temperatures well above $T^*$ [8,9]. In addition, several studies have shown that the itinerant carriers in the PG regime have FL character [5,10,11,12,13], with a carrier density $p$: at low and intermediate hole concentrations, the sheet resistance follows the simple scaling $\rho = A_{2s}T^2/p$ (that can be associated with the Drude expression for the resistivity) for a range of cuprates [5]; the magnetoresistance exhibits Kohler scaling for $T < T^{**}$ [10], a nontrivial characteristic of ordinary single-band metals; the optical scattering rate follows FL temperature/frequency scaling [11]; the Hall constant for $T < T^{**}$ is approximately $T$-independent and a good measure of the doped carrier density ($R_H = 1/(ep)$, where e is the electron charge) [13,14]; and the low-energy effective mass is approximately temperature- and doping-independent in zero magnetic field [11,12]. Based on Hall-angle measurements, however, an even stronger statement can be made: an underlying FL-like transport scattering rate prevails throughout the entire accessible doping range [13,15], connecting the underdoped, strange-metal and overdoped regimes.

These observations reveal a great degree of underlying simplicity, yet it is important to recognize that the cuprates are not ordinary Fermi liquids. In particular, photoemission results show an effectively disconnected Fermi surface ('arcs') in underdoped compounds at low temperatures [16], NMR experiments [17] imply the existence of two local magnetic components – one associated with the PG, and the other with the FL – and specific-heat results are consistent with this scenario [18]. The strong temperature dependence of the effective carrier density, as evidenced by Hall coefficient and Hall angle measurements [13,14], is also highly unusual. Moreover, numerous experiments, such as tunneling [8,19], magnetic resonance [20,21,22], X-ray [23] and neutron [24,25] scattering show that the cuprates are inhomogeneous, both structurally and electronically. In a recent development, a phenomenological model was found to successfully describe both the charge transport and the superfluid density of the cuprates [26]. The model is rooted in two basic experimental facts: an underlying universal transport scattering rate for the itinerant carriers, and a temperature-dependent proliferation of spatially inhomogeneous PGs that commences at temperatures well above $T^*$.

In the present work, our goal is three-fold: (i) to present a detailed resistivity phase diagram of the model cuprate Hg1201; (ii) to test if the data can be quantitatively described by the model of ref. [26], and thereby extend the analysis presented there; (iii) and to investigate how the model captures the similarities and differences among cuprate families, using published charge transport data. Hg1201 is a remarkable cuprate [27]: it is structurally simple, with global tetragonal symmetry, and no known structural phase transitions; it can be hole-doped over a wide range via control of oxygen non-stoichiometry; and it features the highest optimal $T_c$ of any single-$CuO_2$-layer cuprate. Furthermore, the effect of doping-induced point disorder on many observables is weak and similar to the structurally more complex compound $YBa_2Cu_3O_{6+\delta}$ (YBCO), and relatively mild in comparison to, e.g., $La_{2-x}Sr_xCuO_4$ (LSCO) and the Bi-based cuprates. This is evidenced, e.g., by a nearly zero residual resistivity [5,13], the observations of quantum oscillations [28,29] and Kohler scaling [9], the low level of superconducting vortex pinning [5,27], and the clear observation of a vortex lattice in small-angle neutron scattering measurements [30]. Finally, photoemission measurements indicate that the (underlying) Fermi surface is simple near optimal doping [31], in contrast to, e.g., LSCO [32] and YBCO [33]. These properties render Hg1201 a model cuprate system. We find that the temperature-doping resistivity phase diagram of Hg1201 can be comprehensively reproduced by the phenomenological model of ref. [26]. Furthermore, by allowing for small parameter variations, we show that the model provides a quantitative description of the resistivity phase diagrams and the doping dependence of the Hall number of other cuprate families. These findings highlight both underlying universal behavior (e.g., sheet resistance) and expected compound-specific features (e.g., difference in doping dependences of $T^*$). Moreover, they point to a universal structural origin of the inherent gap inhomogeneity that is at best weakly related to doping-induced chemical disorder. It has long been argued that the cuprates are prone to nanoscale structural inhomogeneity, as a means to relax the stresses intrinsic to the perovskite-based structure [34,35]. The role of strong disorder in creating pseudogaps has also been emphasized and related to a broader class of glassy materials [36]. Our results add a fresh emphasis on the need to understand the local structure of the cuprates and its role in determining their unique electronic properties.

The paper is structured as follows: in Sec. II, we present the experimental resistivity phase diagram of Hg1201 and compare the data to the model of ref. [26]; in Sec. III, we analyze published resistivity phase diagrams and recent Hall number data for other cuprate families; in Sec. IV, we discuss the assumptions and implications of the model and analysis. We summarize our findings in Sec. V.

## II. Resistivity phase diagram of Hg1201

We analyze resistivity measurements on twelve Hg1201 single crystal samples, ranging from strongly underdoped to slightly overdoped (Fig. 1a). The dataset is that of ref. [13] with added new measurements for two samples (UD74 and OD90) and a slightly improved estimate of effective sample sizes. The Hg1201 samples were grown and characterized according to established procedures [27,37]. Briefly, the single crystals were grown using an encapsulation method, and annealed in vacuum or oxygen to obtain the desired hole doping level. The doping level is determined from the value of $T_c$ as obtained from Meissner effect measurements [38]. Resistivity is measured with a standard four-contact technique in a Quantum Design, Inc., Physical Property Measurement System (PPMS) using a direct current (with current reversal) and silver-painted contacts to sputtered gold pads on the samples. The contact resistance in all studied samples was on the order of a few Ohm.

Following previous work on other cuprate families [3], we plot the second temperature derivative of the resistivity, i.e., the resistivity curvature, in dependence on temperature and doping in Fig. 1b. The resultant phase diagram shows all the characteristic features present in other cuprates: two temperature scales in the underdoped region of the phase diagram, $T^*$ and $T^{**}$ [5], that correspond to a deviation from high-temperature linear behavior and low-temperature quadratic behavior, respectively; an extended region of approximately $T$-linear resistivity around optimal doping; and a slight positive curvature on the overdoped side of the phase diagram.

All these features can be quantitatively understood with the phenomenological model of [26]. We briefly describe the model here for completeness (for details, see [26]). The model features three premises. First, it uses the essentially universal, Fermi-liquid-like transport scattering rate observed in multiple cuprates [13]: the temperature dependence of the transport scattering rate, as determined from the cotangent of the Hall angle (i.e., the inverse Hall mobility, $\mu^{-1}$), is $\cot(\Theta_H) = \rho/\rho_{xy} = C_2 T^2$, with a universal coefficient $C_2$ throughout the PG, strange-metal and overdoped regimes [15,15]. This strongly suggests that the size of the underlying Fermi surface does not fundamentally change (in agreement with ARPES data [39]), but becomes pseudogapped, and that the scattering process is conventional umklapp electron-electron scattering that does not considerably change as the insulating behavior at zero doping is approached. This experimental fact of a nearly universal scattering rate is a crucial starting point of the model. The second key ingredient of the model is a PG-induced change of the itinerant carrier concentration, with both temperature and doping. Here the main challenge is to connect the underdoped regime, where the low-temperature carrier concentration equals the nominal hole doping level $p$ [14,13], to the overdoped regime, which is known to have a large Fermi surface with $1 + p$ carriers per planar $CuO_2$ unit [6,7]. To make this connection, the model assumes the following. Each $CuO_2$ unit is associated with $1 + p$ holes, with $p$ holes always mobile and one hole separated from the Fermi level by a (doping-dependent) gap $\Delta$. Essentially, this real-space gap corresponds to the $k$-space PG of the underlying large Fermi surface. The third and final premise is that the gap is taken to be inhomogeneous in real space, i.e., to vary from one $CuO_2$ unit to the next. This is quantified through a gap distribution function $G(\Delta)$, and the effective density of itinerant carriers $p_{\text{eff}}$ is obtained in a straightforward manner as the sum of the density $p$ of doped carriers and a temperature- and doping-dependent density of delocalized carriers:

$$p_{eff}(p,T) = p + \int_{-\infty}^{\infty} G(\Delta) e^{-\Delta/2kT} d\Delta \qquad (1)$$

where $k$ is Boltzmann's constant. Physically, the localized states lie below the Fermi level, but we use $\Delta$ with the opposite sign, i.e., it is positive for states below, and negative for states above the Fermi level (to avoid the sign inconsistency between the equations and Fig. 1 in ref. [26]). This convention makes it easier to keep track of the activation terms. For a calculation of $p_{eff}$ (and of all transport coefficients), the distribution function and its doping-dependence need to be specified. The simplest possible assumption, consistent with the approximately linear doping dependence of the PG [40], is that each of the local gaps $\Delta$ decreases linearly with doping and closes eventually. At that point, the hole is no longer localized, and joins the Fermi sea at $T = 0$. The function $G(\Delta)$ will thus contain two parts: a contribution for positive $\Delta$, and a delta-function contribution with a weight equal to the number of unit cells where $\Delta = 0$ at a given doping level. To parametrize the gap distribution, we use a skewed Gaussian of the form $g(\Delta) = 2\Phi(\alpha\tilde{\Delta})\varphi(\tilde{\Delta})$, where $\varphi$ is a normalized Gaussian distribution, $\Phi$ its cumulative (i.e., the error function), $\alpha$ the skew parameter, and $\tilde{\Delta} = (\Delta - \Delta_m)/\delta$, with $\Delta_m$ the Gaussian mean gap and $\delta$ the Gaussian width. This phenomenological parametrization allows a systematic investigation of the influence of distribution width and shape. Using the function $g$, the gap distribution in (1) becomes

$$G(\Delta) = g(\Delta)\theta(\Delta) + \delta(\Delta)\int_{-\infty}^{0} g(\Delta')d\Delta' \qquad (2)$$

where $\theta(\Delta)$ is the Heaviside (step) function and $\delta(\Delta)$ the Dirac delta function. The first term represents the planar $CuO_2$ units with localized holes, while the second term describes those units whose gaps have closed, with the corresponding holes contributing to the Fermi sea at all temperatures. Furthermore, as noted, we assume that the mean gap depends linearly on doping, as $\Delta_m = \Delta_0(1 - p/p_c)$, with $\Delta_0$ the extrapolated mean gap at zero doping, and $p_c$ the crossover doping level where the mean gap is zero. As the simplest possibility, we take the width $\delta$ to be doping-independent. Therefore, four parameters uniquely determine the gap distribution and its doping dependence: $\Delta_0$, $p_c$, $\delta$ and the skew parameter $\alpha$. Importantly, three of the parameters are constrained by various experimental results, as discussed in [25]: $\Delta_0$ is comparable to the charge-transfer gap in the undoped compounds, $\delta$ can be inferred from STM measurements of local gap distributions, and $p_c$ is the doping level where various spectroscopic probes detect the closing of the average pseudogap. In addition, the mid-infrared peak in the optical conductivity follows the mean gap $\Delta$ with doping and has a width comparable to $\delta$. For a Fermi liquid, the (dimensionless) resistivity is simply

$$\rho(p,T) = \frac{C_2}{p_{eff}(p,T)} T^2 \qquad (3)$$

To obtain resistivity values, we use $C_2 = 0.0175$ K$^{-2}$ from experiment [15,15] and multiply with the constant $H/e$ (with $H$ the magnetic field used to obtain $C_2$, which was 9 T for the measurements of Hg1201). Notably, the value of $C_2$ is consistent with conventional electron-electron Umklapp scattering estimates for the large underlying Fermi surface that encloses $1 + p$ carriers [41,42]. The Fermi surface is partially smeared out by the spatially inhomogeneous gaps, yet its remnants are still observed in photoemission experiments at energies below the Fermi level that correspond to the mean gap scale [39].

Figure 1c demonstrates that the model captures the normal-state phase diagram of Hg1201 with remarkable precision up to the highest measured temperature of 400 K. On the underdoped side, the gap distribution is far from the Fermi level, so that at temperatures below $T^{**}$, $p_{\text{eff}} = p$, and the resistivity is quadratic in temperature. The roughly $T$-linear resistivity regime appears when the gap distribution is close to the Fermi level, and holes can be continuously excited across the local gaps with increasing temperature. Beyond optimal doping, all the local gaps eventually close, and the full $1 + p$ Fermi surface is established. Yet this does not occur abruptly, due to the local gap inhomogeneity, but in a continuous manner [13,26]. Importantly, as noted in [26], the results do not critically depend on the shape of the distribution. This is seen from Fig. 2, which shows the calculated phase diagrams and representative resistivities for three different values of the skew parameter. A strong skew pushes the linear-$T$ regime to lower temperatures, but for $p < p_c$, the resistivity is always quadratic in the $T \to 0$ limit. We will return to this point below. The skew might result from, and be amplified by a cooperative effect: since the localization gap derives from electronic correlations among neighboring unit cells, a spatial region without localized carriers could induce a collapse of the gaps in its vicinity. Such an avalanche effect might effectively introduce a low-energy cutoff in the gap distribution, and cause a strong skew around and above optimal doping.

For an additional quantitative comparison between the model and experiment, we extract the low-temperature quadratic resistivity coefficient, $A_2$, in the regime where $\rho = A_2 T^2$ ($T < T^{**}$, red in Fig. 1b,c), and the linear coefficient, $A_1$, for $T > T^*$, similar to previous work [5]. As shown in Fig. 3, the model captures the data well. The doping dependence of $A_2$ is consistent with $A_2 \sim 1/p$ up to at least $p \sim 0.13$, whereas $A_1$ decreases somewhat more strongly with doping. Note also that the model correctly captures the relative magnitudes of $A_1$ and $A_2$, which cannot be separately adjusted within the calculation. Notably, there is no enhancement of either $A_1$ or $A_2$ near optimal doping, neither experimentally [5] nor within the model, inconsistent with the notion of an underlying quantum critical point with fluctuations that affect the itinerant electronic system [43,44,45,46].

### III. Other cuprates

Having demonstrated the success of the phenomenological model from a comparison with Hg1201, we turn to resistivity phase diagrams for other cuprate families from Ref. [3]. Figure 4 shows a comparison for $Bi_2Sr_{2-z}La_zCuO_{6+\delta}$ (BSLCO) and YBCO. A comparison for LSCO was performed in [26]. The gap distribution parameters for the four cuprates are summarized in Table 1. Notably, other cuprates exhibit structural complications that are absent in Hg1201, including orthorhombic distortions (LSCO), superstructure (bismuth-based cuprates), and Cu-O chains (YBCO). Nevertheless, measurements of linear and quadratic resistivity coefficients show a nearly universal evolution with doping [5]. The modeling bears this out, giving good agreement with the respective experimental phase diagrams with quite similar gap distribution parameters. Clearly, the model with a skewed Gaussian gap distribution has enough flexibility to account for material-specific differences, while successfully capturing the universal features of the phase diagram. The gap distribution parameters for the different families are similar, with a somewhat smaller distribution width for YBCO. This points to a universal underlying mechanism of gap inhomogeneity in cuprates, which appears to be inherent to these lamellar perovskite-related oxides and at best weakly related to doping-induced disorder. Namely, LSCO and BSLCO are substitutionally doped, whereas Hg1201 and YBCO are doped with interstitial oxygen, with widely different levels of point disorder affecting the $CuO_2$ planes [47].

In addition to the resistivity data, recent measurements of the Hall number in overdoped cuprates [48] permit a direct comparison with the effective carrier density in our model, under the reasonable assumption that the Hall number is a good measure of carrier density for cuprates with simple Fermi surface geometries [13]. Figure 5 shows the low-temperature Hall number data for $Tl_2Ba_2CuO_{6+\delta}$ (Tl2201) and $Bi_2Sr_2CuO_{6+\delta}$ (Bi2201) from ref. [48], together with the calculated $p_{\text{eff}}$ using gap distribution parameters very similar to the other cuprates (see Table 1). The good agreement supports a key feature of the model – the gradual closing of local gaps with increasing doping, as opposed to a quantum critical point.

## IV. Discussion

The phenomenological model of ref. [26] provides a simple explanation of perhaps the most unusual transport feature of the cuprates, the approximately linear-$T$ behavior near optimal doping from low to high temperatures. This is achieved by forgoing the common assumption that the resistivity temperature dependence is solely determined by the scattering rate, an assumption that is especially far-reaching in the $T$-linear region, since a $T$-linear scattering rate is often interpreted as indicative of quantum critical fluctuations [1,46]. Yet this is clearly incompatible with the experimentally-determined underlying universal Hall mobility [15,15]. Notably, the Hall mobility includes a ratio of the transport lifetime and effective mass and, in principle, it is possible that their doping and temperature dependences conspire to give universal $T^2$ behavior. This is highly unlikely, however, as it would require a remarkable level of fine-tuning across the whole phase diagram [13,15]. Notably, optical experiments suggest a doping-independent effective Drude mass in several cuprates [12]. Furthermore, exquisite measurements of the superfluid density of LSCO do not show any anomalies that could be associated with an increased zero-field effective mass of the superconducting carriers around $p \sim 0.2$ [49]. In contrast, quantum oscillation studies of the Fermi surface in high magnetic fields show evidence for effective mass changes with doping [50]. Yet it seems likely that the high-field state with a reconstructed Fermi surface is related to CDW order and qualitatively different from the zero-field normal state [29], since there are no visible CDW-related features in the zero- and low-field resistivity phase diagrams (see also below). A recent specific-heat study of co-doped La-based cuprates argues for an increased Sommerfeld coefficient, and therefore an increased carrier effective mass, around a critical doping [43]. However, both the itinerant and the localized subsystem should contribute to the specific heat. Regarding the itinerant subsystem, one should be careful about the presence of a Van Hove singularity at the Fermi level in La-based cuprates. One would also expect an increase of the low-temperature effective Sommerfeld coefficient in the doping range where the localization gap structure crosses the Fermi level, since holes can be continuously thermally excited across the local gaps.

If the scattering rate is simply quadratic in temperature and the effective mass essentially constant, the nontrivial behavior of the resistivity is due to a temperature-dependent carrier density, as indeed supported by the modeling. Notably, this argument has been used in previous work to model the transport coefficients of underdoped cuprates [14,51,52], and it is consistent with Hall-effect measurements [13,14,53,54]. We emphasize though that the scattering rate in our model remains quadratic throughout the phase diagram, and in the limit of zero temperature, for the resistivity is quadratic as well. This is in stark contrast to the quantum phase transition scenario [1,43,46], in which the linear-$T$ resistivity originates from scattering off quantum fluctuations near a critical point. In this scenario, the zero-temperature limit of the resistivity is linear, with possible deviations at higher temperatures and away from the putative quantum critical doping level [45]; furthermore, the anomalous scattering mechanism should cause the coefficients $A_1$ and $A_2$ to peak around the critical doping, which is not observed for Hg1201 (Fig. 3) or other cuprates [5]. In the cuprates, it is notoriously difficult to determine the true low-temperature normal-state behavior, because of the extremely high magnetic fields needed to completely suppress superconductivity and the possibility of a field-induced modification of the normal state [4,55]. For compounds with lower critical fields, such as Nd-doped LSCO, there have been reports of low-temperature linear-$T$ resistivity and constant Hall mobility [44]; yet we emphasize that these compounds are structurally and electronically complex, with several low-temperature structural transitions that involve soft phonons [56], structural instabilities [57] and high residual resistivities [3,44]. Furthermore, several cuprates (including the La-based materials) feature a van Hove singularity in the density of states above optimal doping that should affect the low-temperature transport properties [58] and possibly the specific heat [43]. Importantly, however, our focus is on the gross features of the phase diagram, up to the comparatively high temperature/energy scale relevant to understand the high-$T_c$ phenomenon, that can clearly be understood without invoking quantum criticality. We also note that, for YBCO [3] and Hg1201 (Fig. 1a), the low-temperature resistivity cannot be linear in temperature (with the same slope as at high temperature) around optimal doping. Namely, if one extrapolates the linear dependence above $T_c$ to $T = 0$, one finds a *negative* residual resistivity, which implies that the underlying normal-state resistivity must have curvature at low temperatures. Furthermore, both YBCO and Hg1201 feature relatively small residual resistivities [3,5] due to the relatively gentle effects of oxygen doping [47], which has enabled the observation of quantum oscillations in underdoped samples [28,29,59]. Yet the same argument cannot be made for LSCO or BSLCO, where point disorder due to substitutional doping induces large residual resistivities at all doping levels [3,4] and resistivity upturns for underdoped samples [3,15,60]. Importantly, in the doping range of the present study, there is no evidence for a doping-dependent zero-field effective mass, and hence the linear resistivity coefficient is inconsistent with the recent assertion of a universal 'Planckian dissipation limit' [61]. This is the case not only for Hg1201, but also for the other cuprates analyzed here.

Another important point is the absolute value of the resistivity, which can be fairly high compared to 'good' metals. Although it is often suggested that the cuprates exceed the semi-classical Mott-Ioffe-Regel limit for coherent charge transport, arguments have been put forward that the relevant limit is in fact significantly higher due to electronic correlations [5,13,62]. The high resistivity values at moderate and low doping are mainly the result of a low charge density (with an underlying large Fermi surface), whereas the room-temperature Hall mobility of the cuprates such as Hg1201 is comparable to that of ordinary metals such as Aluminum [13]. We note that there is indeed some experimental evidence for an approach to resistivity saturation, e.g., in strongly underdoped LSCO at high temperatures [2,62]. Naturally, at high enough temperatures, the premises of our model will break down, in at least two possible ways: (i) additional scattering mechanisms such as coupling to optic phonons will modify the simple $T^2$ scattering rate; (ii) the resistivity values will approach saturation, yet at a significantly higher level than the semi-classical Mott-Ioffe-Regel limit [62]. However, since the available Hall mobility data up to 400 K show no evidence for these deviations, there is no need to include them in the modeling.

The resistivity phase diagram is rather insensitive to the different electronic ordering phenomena in underdoped cuprates, most importantly weak charge-density-wave (CDW) order [1]. Quasi-static two-dimensional CDW correlations with short coherence lengths and small amplitudes have been mapped as a function of doping and temperature in, e.g., in YBCO [63,64] and Hg1201 [65,66,67] via resonant X-ray scattering and pump-probe optical experiments, and found to be strongest in the underdoped part of the phase diagram (around $p \sim 0.12$ and $p \sim 0.09$, respectively, for YBCO and Hg1201). In addition, dynamic CDW correlations are present in a wide doping/temperature range [68,69]. Yet no particular features related to the characteristic CDW temperatures are seen in the normal-state resistivity phase diagrams considered in the present work, and only slight downturns are visible in the low-field Hall constant above $T_c$ around $p \sim 0.1$ [13] (compared to the substantial effect in high-field Hall data [70] at temperatures below the zero-field $T_c$ that is associated with a reconstructed Fermi surface). This indicates that the CDW-induced Fermi-surface reconstruction [29,65,71] is a secondary phenomenon that only occurs in high magnetic fields and at temperatures below the zero-field $T_c$.

Notably, the mean gap scale $\Delta_m$ is also seen in photoemission and tunneling experiments [26,40], and it manifests itself as a broad peak in the optical conductivity, detected in several representative cuprate families [13,26,72,73]. It is the highest intrinsic gap scale in a hierarchy of (pseudo)gaps observed by different experimental probes in underdoped cuprates [26,40]. While firmly rooted in experiment, the model tested here is phenomenological in nature and does not give microscopic insight into the origin of the local gap, its doping dependence, or the gap distribution function. The model uses the simplest possible assumptions for the mean gap and distribution width doping dependences, which in reality may be somewhat different, especially for strongly underdoped compounds close to the insulating phase [26]. The inhomogeneous gap can be interpreted as a localization gap, induced by strong electronic correlations. Accordingly, the extrapolated mean gap at zero doping is within a factor of two of the transport charge-transfer gap [14]; this discrepancy can be remedied if a nonlinear gap dependence on doping is used [26]. The decrease of the localization gap with doping must be due to screening, and the model could thus be refined by obtaining the doping dependence of the gap distribution in a self-consistent way. Notably, the localization gap scale decreases relatively slowly with doping, compared to simple Mott insulators such as doped $LaTiO_3$. This is most likely a consequence of the two-component electronic subsystem, which might be viewed as orbital-selective Mott physics [74,75].

The good agreement with the experimental resistivity phase diagrams, using a doping-independent gap distribution width, is consistent with the finding of gap disorder in STM experiments [8,19,76,77,78]. This intrinsic inhomogeneity may be of electronic origin; e.g. it has long been known that doped Mott insulators are prone to phase separation and spatial modulation of charge [79,80]. However, it is a distinct possibility that the inhomogenity has an inherently structural origin. Namely, materials with perovskite or perovskite-derived structure are known to be prone to both long- and short-range structural instabilities and texturing [34,36,81]. The basic structural building blocks of the cuprates, the copper-oxygen octahedra (or tetrahedra in some compounds), can distort in multiple ways, which generally leads to low-symmetry average structure in most cuprates, Hg1201 being a notable exception. Yet strain accommodation can also lead to short-range correlated structural disorder, i.e., bond-length and angle modulations, even in crystals with high average symmetry [82]. Due to the long-range nature of elastic forces, such correlations can span multiple length scales, leading to intricate structural features [34]. The bond angles and lengths couple directly to the parameters of the electronic Hamiltonian, and hence local structural distortions should be expected to profoundly influence the electronic system in a nonlinear fashion. The effects of such intrinsic inhomogeneity on superconducting fluctuations have recently been documented in cuprates through resistivity, nonlinear response, and torque magnetization experiments, showing a universal percolative regime [83,84,85]. The disappearance of superconductivity at high doping in LSCO is also consistent with a percolative scenario [26]. Similar superconducting fluctuation physics was also found more broadly in perovskite-based materials [86], which demonstrates a profound relation between structure and electronic inhomogeneity. It is hence to be expected that the normal state in the cuprates is affected as well. Through a coupling of localized holes to the lattice, subtle inhomogeneity in bond angles and distances can induce a localization gap distribution, which could also be amplified by the tendency of the electronic system towards inhomogeneity. This conclusion is supported by the observations that the tetragonal-to-orthorhombic structural transition in LSCO [87] corresponds to ~100% localization of one hole per unit cell [26], and of a subtle breaking of inversion symmetry in YBCO in a similar temperature/doping range [88]. Importantly, the dichotomy between the localized and itinerant subsystems is reflected in a sensitivity of the antinodal states to inhomogeneity, whereas the itinerant, nodal states are largely unaffected. This is an interesting and unusual feature of the cuprates, and its understanding will likely underpin the microscopic picture of our model.

**V. Summary**

To conclude, we have presented a comprehensive resistivity curvature phase diagram for the simple tetragonal cuprate Hg1201 and used this to test a recent phenomenological model for the normal state in a quantitative fashion. The model employs a renormalized gap scale that is linked to the charge-transfer gap of the undoped insulating state, and that constitutes the largest pseudogap scale at non-zero doping. Importantly, one of the key assumptions of the model – that local (pseudo)gaps close at different temperatures and doping levels – is upheld in a comparison to Hall number data. This seems to rule out a quantum phase transition scenario for the cuprate phase diagram, where the Fermi surface would abruptly change at a critical doping level. Instead, the change in carrier density is gradual, both upon doping and with increasing temperature, due to the inhomogeneous local gaps. Furthermore, we have demonstrated that the model is remarkably successful in reproducing the phase diagrams of other cuprate families as well, with nearly universal parameters. This universality implies that the considerable gap distribution is an inherent characteristic of the $CuO_2$ planes, and quite insensitive to the details of the crystal structure or doping method. However, the differences in model parameters for the different cuprate families (in particular the distribution widths) could provide important clues to the precise origin of the inhomogeneity, when combined with experiments that probe the local electronic structure.

**Acknowledgments.**

We thank M. K. Chan for contributions to the crystal growth and transport measurements. The work at the University of Minnesota was funded by the Department of Energy through the University of Minnesota Center for Quantum Materials under DE-SC-0016371. The work at the TU Wien was supported by the European Research Council (ERC Consolidator Grant No 725521), while the work at the University of Zagreb by project CeNIKS co-financed by the Croatian Government and the European Union through the European Regional Development Fund - Competitiveness and Cohesion Operational Programme (Grant No. KK.01.1.1.02.0013) and the Croatian-Swiss Research Program of the Croatian Science Foundation and the Swiss National Science Foundation with funds obtained from the Swiss-Croatian Cooperation Programme.

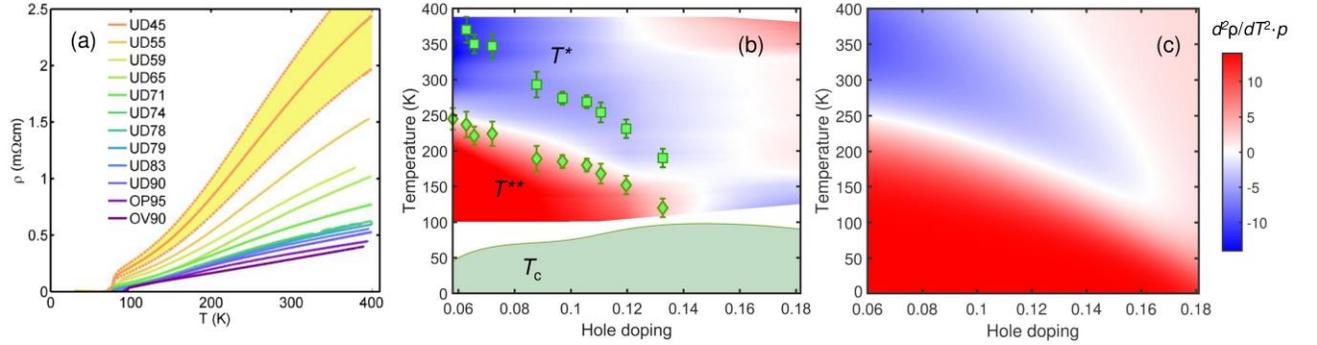

Figure 1. Resistivity phase diagram of Hg1201. (a) Raw planar resistivity data for 12 single crystal samples of Hg1201, marked by their $T_c$ values; UD, OP and OV represent underdoped, optimally doped, and overdoped samples, respectively. The band around the UD45 curve indicates the typical (relative) uncertainty of the absolute resistivity value due to crystal dimensions and contact distance. Data for UD45 – UD71 and UD78 – OP95 are from ref. [13], and the data for UD74 and OV90 are new. (b) Experimental and (c) calculated contour plot of the second temperature derivative (curvature) of the resistivity. The experimental plot is obtained from data in (a). The resistivity curvature is normalized by the hole doping $p$, which is obtained from $T_c$ values [38], and the color scale is the same for (b) and (c). The symbols in (b) are the characteristic temperatures $T^*$ and $T^{**}$ obtained from the data: $T^*$ is the temperature below which the resistivity curves depart from high-temperature linear behavior, whereas $T^{**}$ indicates the departure from low-temperature quadratic behavior. The model parameters used to obtain (c) are listed in Table 1.

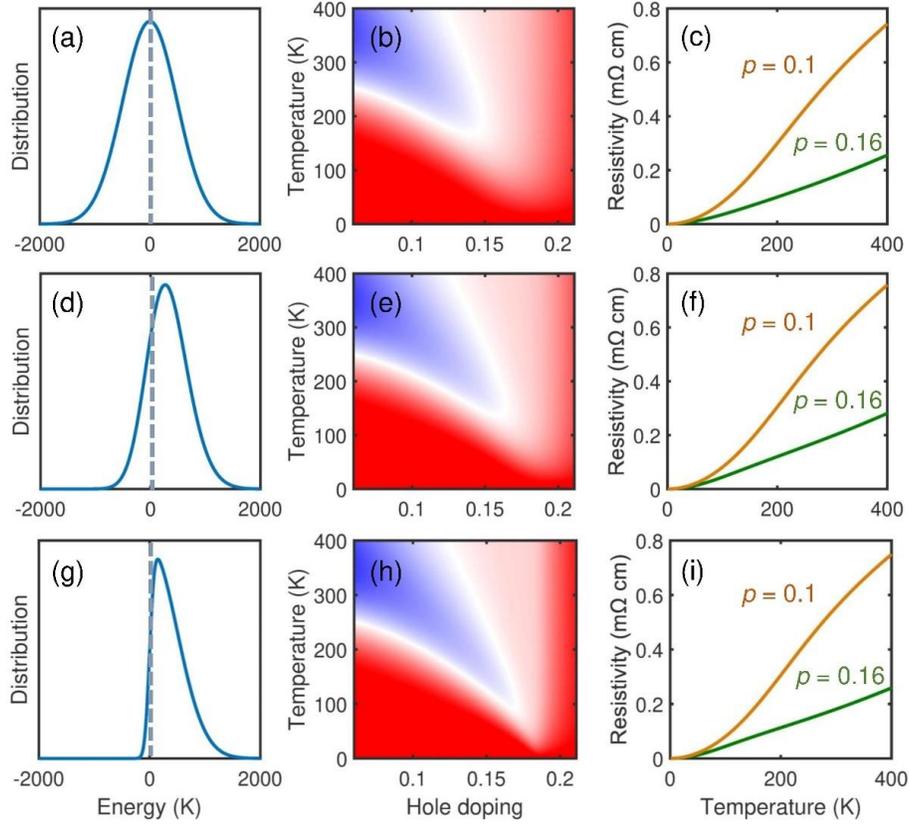

Figure 2. Modeling of planar resistivity. Gap distribution functions, resistivity curvature phase diagrams and representative resistivity curves for three different values of the distribution skew parameter: (a)-(c), α = 0 (pure Gaussian distribution); (d)-(f), α = 2 (typically used for comparison with experiment); (g)-(i), α = 10 (strong skew). The distributions are shown for $p = p_c$. The crossover doping levels were adjusted for each distribution to yield the most similar curves at $p = 0.16$, and were $p_c = 0.22$ in (a)-(c), $p_c = 0.195$ in (d)-(f) and $p_c = 0.18$ in (g)-(i). In all panels, the gap distribution width is $\delta = 700$ K and $\Delta_0 = 4000$ K

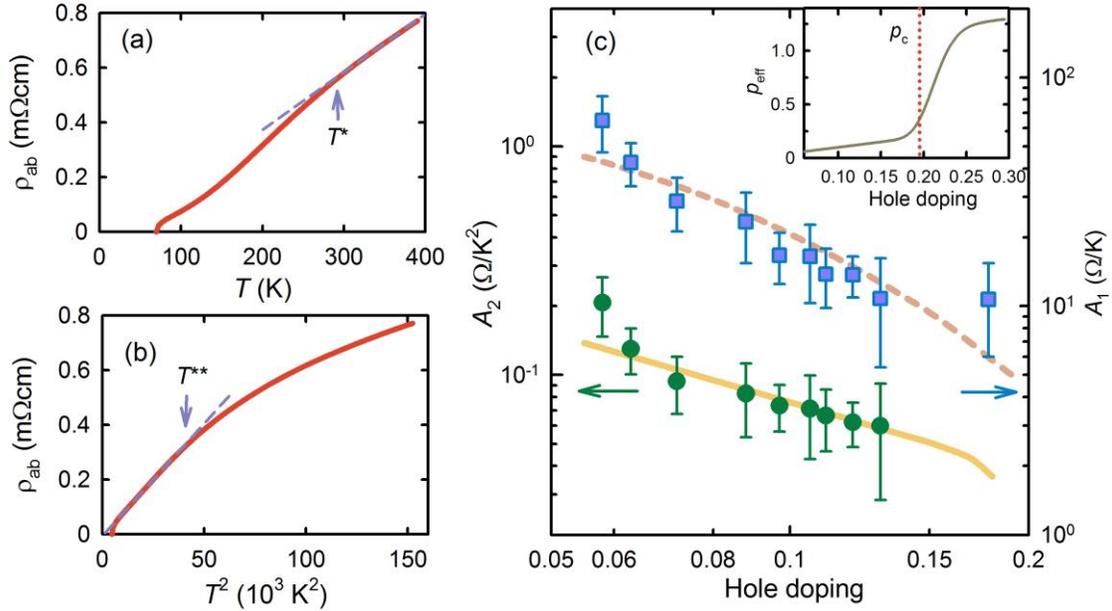

Figure 3. Resistivity coefficients of Hg1201. Resistivity of the UD71 sample on (a) linear and (b) quadratic temperature scales, with characteristic temperatures $T^*$ and $T^{**}$ and linear and quadratic slopes (coefficients) $A_1$ and $A_2$. (c) Linear (circles) and quadratic (squares) resistivity coefficients of Hg1201, obtained by fitting the data above $T^*$ and below $T^{**}$, respectively, as in (a) and (b). The lines are obtained from the model, with the same parameters as in Fig. 1c. The quadratic coefficient is consistent with $A_2 \sim 1/p$ up to about optimal doping. Note that, in the model, $A_2 \sim 1/p_{\text{eff}}(p,0)$, since the low-temperature asymptotic behavior is always quadratic in temperature. The model yields $A_1$ without free parameters once $A_2$ is given, and thus correctly captures the absolute value of the ratio $A_1/A_2$. The inset gives $p_{\text{eff}}(p,0)$ from Eq. (1) in a wide doping range and demonstrates the smooth change around $p_c$.

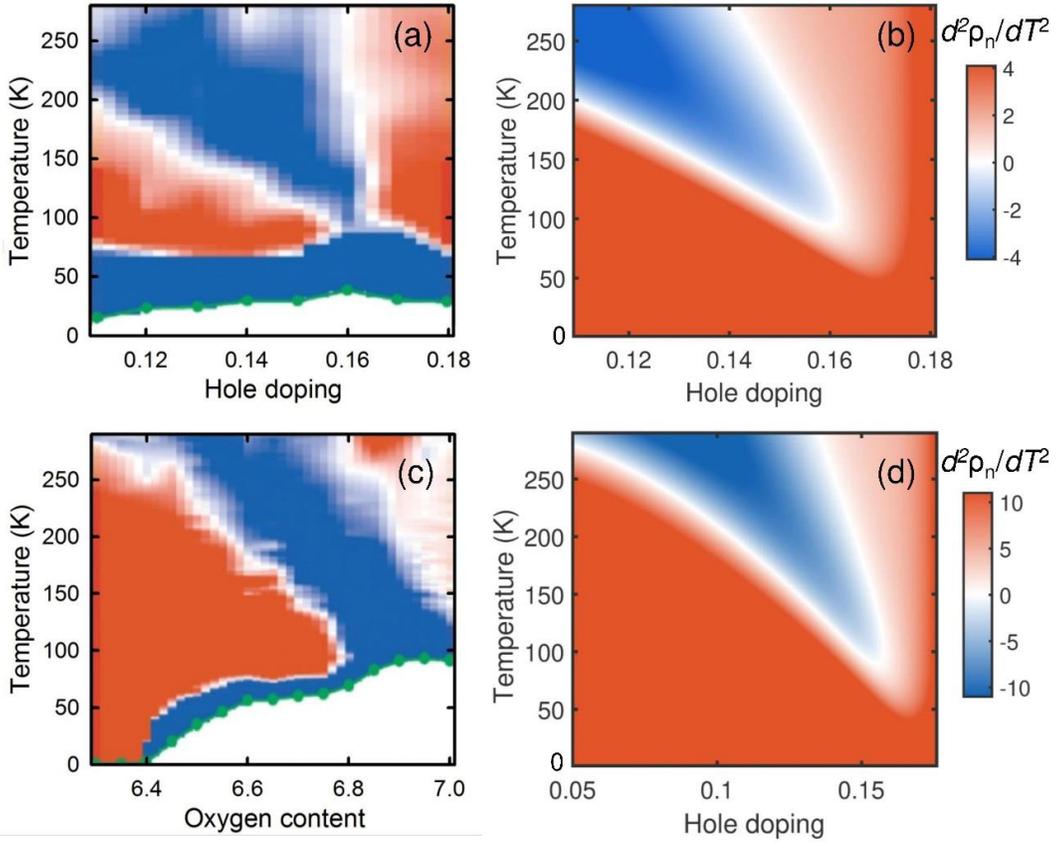

Figure 4. Resistivity phase diagrams of other cuprate families. (a) Experimental and (b) model resistivity phase diagram for BSLCO, and (c) experimental and (d) model phase diagram for YBCO. The data are from ref. [3], and the model parameters are listed in Table 1. Experimental resistivity curves are normalized to the values at 280 K (BSLCO) and 290 K (YBCO); calculated curvatures are normalized by $p_{\text{eff}}(p,0)$. The experimental phase diagrams include superconducting domes and the corresponding fluctuation regimes, while the model phase diagrams are limited to the normal state.

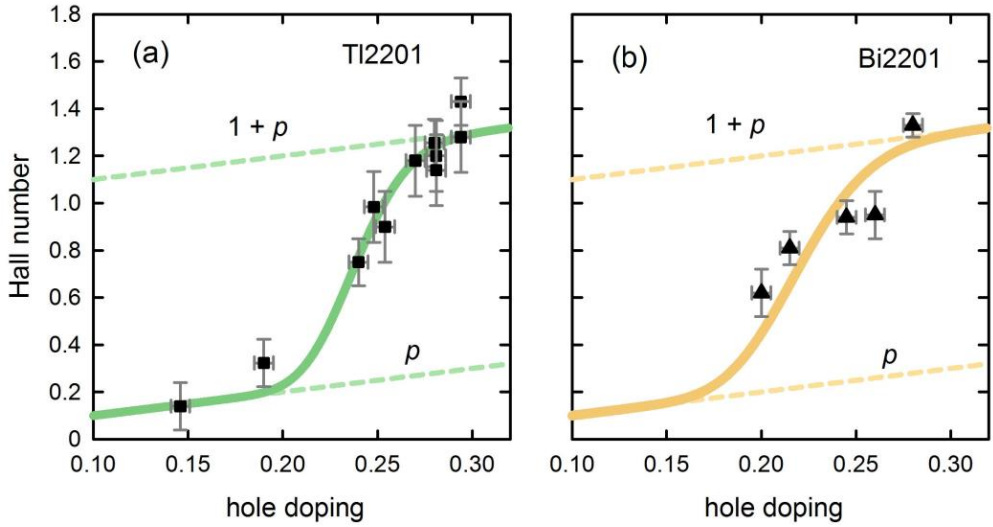

Figure 5. Doping dependence of the carrier number for (a) Tl2201 and (b) Bi2201, demonstrating the smooth crossover from $p$ to $1 + p$ carriers across optimal doping. Experimental data are from ref. [48], and the curves are obtained from the model using the estimated parameters listed in Table 1. Note that the curves are not fits (since especially in the case of Bi2201 the fitting procedure would be unreliable), and a variation of roughly ±10% in the gap distribution parameters would not change the agreement much.

|        | $\Delta_0$ | $\delta$ | $p_c$   | $\alpha$ |
|--------|------------|----------|---------|----------|
| Hg1201 | 3800 K     | 700 K    | 0.197   | 2        |
| LSCO   | 3900 K     | 800 K    | 0.22    | 2        |
| BSLCO  | 4300 K     | 700 K    | 0.175   | 4        |
| YBCO   | 4500 K     | 500 K    | 0.177   | 2        |
| Tl2201 | ~3700 K    | ~700 K   | ~0.22   | 2        |
| Bi2201 | ~3500 K    | ~1000 K  | ~0.2    | 2        |

Table 1. Gap distribution parameters for six cuprate families. The parameters for LSCO are from ref. [26], and the parameters for Tl2201 and Bi2201 are rough estimates obtained from comparison with Hall number data only (Fig. 5)